\documentstyle[12pt]{article}
\newcommand{\beq}{\begin{eqnarray}}
\newcommand{\eeq}{\end{eqnarray}}
\begin{document}
\title{Upsilon Production In p-p Collisions at LHC}
\author{Leonard S. Kisslinger\\
Department of Physics, Carnegie Mellon University, Pittsburgh PA15213}
\date{}
\maketitle
\noindent
PACS Indices:12.38.Aw,13.60.Le,14.40.Lb,14.40Nd
\vspace{1mm}

\begin{abstract}
  This is a continuation of recent studies of $\Upsilon(nS)$ production at 
the LHC in p-p collisions. Our previous studies were for 2.76 TeV, while
the present predictions are for 7.0 TeV collisions. 
\end{abstract}
\vspace{3mm}

 In our recent work\cite{kmm11} on heavy quark state production in p-p
(proton-proton) collisions we calculated $\Upsilon(nS)$ production using
the color octet model, which was shown to dominate color singlet production
of heavy quark states in p-p collisions\cite{nlc03,cln04}. We use the
treatment formulated by Nayak and Smith\cite{ns06}. The main
objective of this work is to study how our mixed hybrid theory differs
from the standard quark model for $\Upsilon(nS)$ production. In our 
theory\cite{lsk09} the $\Upsilon(1S)$ and $\Upsilon(2S)$ states are
conventional $b\bar{b}$ states, while the $\Upsilon(3S)$ is 50\% standard
$b\bar{b}$ and 50\% hybrid. See Ref\cite{kmm11} for a brief sumary of this
theory. For the LHC production we used an energy of 2.76 TeV, corresponding 
to the publication of CMS results, mainly for Pb-Pb collisions\cite{cms11},
and calculated the ratios of cross sections $ \frac{\sigma(\Upsilon(2S))+
\sigma(\Upsilon(3S))}{\sigma(\Upsilon(1S))}$, since the 2S and 3S states
were not resolved for p-p collisions in Ref\cite{cms11}. In light of new 
results by the CMS Collaboration for p-p collisions at at 
7.0 Tev\cite{cmsprd11}, with a much larger data set, in which both the 
$\Upsilon(2S)$ and $\Upsilon(3S)$ ratios to the $\Upsilon(1S)$ cross 
sections were measured, we submit this note.

  Although we use the color octet model, we note that there have been
many publications on the color singlet vs the color octet
model. For details of the color singlet model see the review by 
Lansberg\cite{lans06} in which the color singlet model for $J/\Psi$ 
and $\Upsilon$ production is found to be much larger than the color singlet 
model. This approach, however, involves complicated calculations 
with many higher order diagrams. For example, Lansberg $et$ $al$\cite{lck06}
found higher order processes which are color-octet-like to be as large as 
color singlet ones. As stated above, in Refs\cite{nlc03,cln04} the color 
octet processes were found to be much larger than the color singlet for
heavy quark meson production. Also, it should be noted that gluonic 
processes are dominant for both singlet and octet models, and our results 
are based on the hybrid nature of certain heavy quark states, with valence 
gluons\cite{kmm11}. Since we only find ratios of matrix elements, our 
mixed hybrid model results might be similar for the singlet model as the 
octet model, an interesting topic for future work.

  We now briefly review the theory. Using the color octet model with 
scenerio 2 (see Refs\cite{kmm11,ns06}), the cross section for helicity 
$\lambda=0$, which is dominant, is
\beq
\label{1}
  \sigma_{pp\rightarrow \Upsilon(\lambda=0)} &=& A_\upsilon \int_a^1 
\frac{d x}{x} f_g(x,2m)f_g(a/x,2m) \; , 
\eeq
where $f_g(x,2m)$ are the gluonic distributions evaluated at 2m=10GeV,
corresponding to the bottom quark mass. The quantities $a= 4m^2/s$ and
$A_\Upsilon=\frac{5 \pi^3 \alpha_s^2}{288 m^3 s}<O_8^\Upsilon(^1S_0)>$ in
scenerio 2. Note that the quantity $a$ decreases with energy, increasing
the cross section, while the quantity $A_\Upsilon$ decreases with energy, 
with a net result that in the
octet model the cross section decreases with energy. On the other hand,
the cross cections depend on the scenerios\cite{kmm11}, and the only
meaningful calculations in the present work are in the ratios of cross 
sections.

  As derived in Ref\cite{kmm11}, in the standard model, using harmonic
oscillator wave functions, since with bottom quarks nonrelativistic
theory is adequate, we find for the ratios of cross sections at 7.0 TeV
in the standard model
\beq
\label{2S3S/1S}
      \sigma(\Upsilon(2S))/\sigma(\Upsilon(1S))&\simeq& 0.27 {\rm \;\;standard}
\nonumber \\
      \sigma(\Upsilon(3S))/\sigma(\Upsilon(1S))&\simeq& 0.04 {\rm \;\;standard}
 \; ,
\eeq
while in the mixed hybrid picture the ratio 
$\sigma(\Upsilon(2S))/\sigma(\Upsilon(1S))$ is 0.27 as in the standard model,
while (see Ref\cite{kmm11} for hybrid vs standard for $\Upsilon(3S)$ 
production) 
\beq
\label{3S/1S}
  \sigma(\Upsilon(3S))/\sigma(\Upsilon(1S))&\simeq& 0.1{\rm \;mixed\;hybrid} 
\; .
\eeq 

The CMS results at 7.0 TeV are\cite{cmsprd11}
\beq
\label{CMS11}
 \sigma(\Upsilon(2S))/\sigma(\Upsilon(1S))&\simeq& 0.26 \pm 0.02 \pm 0.04
  {\rm \;\;CMS} \nonumber \\
  \sigma(\Upsilon(3S))/\sigma(\Upsilon(1S))&\simeq& 0.14 \pm0.01 \pm 0.02
 {\rm \;\;CMS} \; .
\eeq

  Therefore one sees that the CMS results for the 
$\sigma(\Upsilon(3S))/\sigma(\Upsilon(1S))$ ratio are in disagreement with
the standard quark model, but agree within errors with the mixed hybrid
theory. Note that new CMS results for $\Upsilon(nS)$ cross sections with
much more data are to be published soon\cite{cms12}, and are in agreement
with the data of Ref\cite{cmsprd11}.
\vspace{5mm}

\Large{{\bf Acknowledgements}}\\
\normalsize
This work was supported in part by a grant from the Pittsburgh Foundation.
The author thanks Professor Thomas Ferguson for information on CMS results
and helpful discussions.

\end{document}